\documentclass[12pt,reqno]{amsart}
\usepackage{amsfonts}
\usepackage{mathrsfs}
\usepackage{amsmath}
\usepackage{amssymb,amsfonts}
\usepackage{cite}

\newtheorem{thm}{Theorem}[section]
\newtheorem{lem}{Lemma}[section]

\numberwithin{equation}{section}

\def\pf{{\textit {Proof:} }}

\newcommand{\mysection}[1]{\section{#1}\setcounter{equation}{0}}

\newfont{\bb}{msbm10 at 11pt}

\newcommand{\bal}{\begin{aligned}}      \newcommand{\eal}{\end{aligned}}
\newcommand{\ba}{\begin{array}}      \newcommand{\ea}{\end{array}}
\newcommand{\bc}{\begin{center}}     \newcommand{\ec}{\end{center}}
\newcommand{\be}{\begin{enumerate}}  \newcommand{\ee}{\end{enumerate}}
\newcommand{\beq}{\begin{eqnarray}}  \newcommand{\eeq}{\end{eqnarray}}
\newcommand{\beQ}{\begin{eqnarray*}} \newcommand{\eeQ}{\end{eqnarray*}}
\newcommand{\bi}{\begin{itemize}}    \newcommand{\ei}{\end{itemize}}
\newcommand{\bt}{\begin{tabular}}    \newcommand{\et}{\end{tabular}}
\newcommand{\bdm}{\begin{displaymath}} \newcommand{\edm}{\end{displaymath}}
\newcommand{\pt}{\partial}

\newcommand{\pr}{\partial r}
\newcommand{\rw}{\rightarrow}

\newcommand{\ep}{\epsilon}
\newcommand{\ir}{\int_{0}^{r}}

\newcommand{\Ba}{\Big |}

\newcommand{\ift}{\infty}

\newcommand{\Bl}{\Big (}

\newcommand{\bhn}{\bar{h}_n}
\newcommand{\bgn}{\bar{g}_n}

\newcommand{\Br}{\Big )}
\newcommand{\ur}{\left(1+u+r\right)}
\newcommand{\uu}{\left(1+u\right)}

\newcommand{\cn}{\chi_n}

\newcommand{\hy}{\|h_n-h_{n-1}\|_Y}

\def\qed{\hfill{Q.E.D.}\smallskip}

\newcommand{\ls}{\setlength{\baselineskip}{12pt}
                 \setlength{\parskip}{3mm}}

\begin{document}

\allowdisplaybreaks

\title[Einstein-scalar-field equations]{Spherically symmetric Einstein-scalar-field equations for slowly particle-like decaying null infinity}

\author[C Liu]{Chuxiao Liu$^{1,4}$}
\author[X Zhang]{Xiao Zhang$^{2,3,4}$}

\address[]{$^{1}$School of Mathematics and Information Science, Guangxi University, Nanning, Guangxi 530004, PR China}
\address[]{$^{2}$State Key Laboratory of Mathematical Sciences, Academy of Mathematics and Systems Science, Chinese Academy of Sciences, Beijing 100190, PR China}
\address[]{$^{3}$School of Mathematical Sciences, University of Chinese Academy of Sciences, Beijing 100049, PR China}
\address[]{$^{4}$Guangxi Center for Mathematical Research, Guangxi University, Nanning, Guangxi 530004, PR China}

\email{cxliu@gxu.edu.cn$^{1,4}$}
\email{xzhang@amss.ac.cn$^{2,3,4}$}


\date{}

\begin{abstract}
We show that the spherically symmetric Einstein-scalar-field equations for small slowly particle-like decaying initial data at null infinity have unique global solutions.
\end{abstract}

\maketitle \pagenumbering{arabic}

\mysection{Introduction}
\ls

Spherically symmetric spacetime metrics can be written as
\beq
ds^2=-gqdu^2-2gdudr+r^2(d\theta ^2+\sin^2 \theta d\psi^2)\label{me}
\eeq
in Bondi coordinates, see, e.g. \cite{C1, LZ, LZ2}, where $g(u,r)$ and $q(u,r)$ are $C^2$ nonnegative functions over $(0,\infty)$. The null frames are
\beQ
\vec n=\frac{1}{\sqrt{gq}}D,\quad \vec l=\frac{1}{\sqrt{gq^{-1}}}\frac{\pt}{\pt r},\quad e_1=\frac{1}{r}\frac{\partial }{\partial \theta}, \quad
    e_2=\frac{1}{r \sin \theta }\frac{\partial }{\partial \psi},
\eeQ
where
\beQ
D=\frac{\pt}{\pt u}-\frac{q}{2}\frac{\pt }{\pt r}
\eeQ
is the derivative along the incoming light rays.

Throughout the paper, we denote by $\bar f(r)$ the integral average of integrable function $f(r)$ over $[0, r]$
\beQ
\bar f(r)=\frac{1}{r}\int_{0}^{r} f(r')dr'.
\eeQ

For a real $C^2$ scalar field $\phi(u,r)$ on $(0, \infty)\times (0, \infty)$,
the Einstein-scalar field equations are
\begin{align}
R_{\mu\nu}=8\pi\partial_{\mu}\phi\partial_{\nu}\phi, \quad \Box{\phi} =0.  \label{einstein-scalar}
\end{align}
Under the following regularity conditions at $r=0$ and boundary condition at $r=\infty$,

\noindent{\bf Regularity Condition I}: For each $u$,
\begin{align*}
\lim _{r \rightarrow 0} \left (- \frac{rq}{g} \frac{\partial g}{\partial u} +r \frac{\partial q}{\partial u} +\frac{rq^2}{2g} \frac{\partial g}{\partial r}
-8\pi r^2 \Big(\frac{\partial \phi}{\partial u}-\frac{q}{2} \frac{\partial \phi}{\partial r} \Big) ^2 \right)=0. 
\end{align*}

\noindent{\bf Regularity Condition II}: For each $u$,
\begin{align*}
\lim _{r \rightarrow 0} \big(r \phi \big)=\lim _{r \rightarrow 0} \big(r q \big)=0. 
\end{align*}

\noindent{\bf Boundary Condition}: For each $u$,
\begin{align*}
\lim _{r \rightarrow \infty} g=\lim _{r \rightarrow \infty} q=1, 
\end{align*}
the Einstein-scalar field equations \eqref{einstein-scalar} are equivalent to the following systems, see, e.g. \cite{LZ2}
\beq
\begin{matrix}
\left\{\begin{aligned}
h&=\frac{\partial (r\phi)}{\partial r},\\
g&=\exp{\left\{-4\pi\int_{r}^{\infty}{\frac{(h-\bar h)^2}{r'}dr'}\right\}},\\
q&=\bar g=\frac{1}{r}\int_{0}^{r}{gdr'},\\
Dh&=\frac{g-\bar g}{2r}(h-\bar h).
\end{aligned}
\right.\label{D}
\end{matrix}
\eeq

The Bondi mass $M_B(u)$ for each $u$ and the final Bondi mass $M_{B1}$ are given by \cite{BBM, LZ}
\beQ
M_B(u)=\lim _{r \rightarrow \infty} \frac{r}{2}\Big(1-q\Big), \quad M _{B1}=\lim _{u \rightarrow \infty} M_B(u).
\eeQ
The Bondi-Christodoulou mass $M(u)$ for each $u$ and the final Bondi-Christodoulou mass $M_1$ are given by \cite{C1, C4, LZ}
\beQ
M(u)=\lim _{r \rightarrow \infty} \frac{r}{2}\Big(1-\frac{q}{g}\Big), \quad M _1=\lim _{u \rightarrow \infty} M(u).
\eeQ

In \cite{C1,C2,C3}, Christodoulou studied global existence and uniqueness of classical solutions with small initial data, and of generalized solutions with large initial data, when the initial data satisfies the following decaying property
\begin{align*}
h(0,r)=O(r^{-3}),\quad \frac{\partial h}{\partial r}(0,r)=O(r^{-4}),\quad as\,\,r\rightarrow \infty.
\end{align*}
In particular, he proved that the solution satisfies the following uniformly decaying estimates
\beQ
\big| h(u,r) \big| \leq \frac{C}{(1+u+r)^{3}}, \quad \Big| \frac{\partial h}{\partial r}(u,r)\Big| \leq \frac{C}{(1+u+r)^{4}},
\eeQ
and the corresponding spacetime is future causally geodesically complete with vanishing final Bondi-Christodoulou mass.

For the Einstein-scalar field equations in double null coordinates
\begin{align}
ds^2=-\Omega ^2 du dv +r^2 \big(d\theta^2+\sin^2\theta d\phi^2 \big), \label{double-null}
\end{align}
Christodoulou also showed the unique spherically symmetric global solution exists for the characteristic initial-value problem with small bounded variation norms \cite{C5}. Later, his result was extended to the more general situation by Luk, Oh and Yang, and they showed the unique spherically symmetric global solution exists and the resulting spacetime is future causally geodesically complete if the initial data satisfies
\beQ
\begin{aligned}
\int _u ^v \big| \Phi(v') \big| dv'  \leq \epsilon \big(v-u \big)^{1-\gamma}, \quad
|\Phi(v)| + \Big| \frac{\partial \Phi} {\partial v} (v)\Big| \leq \delta
\end{aligned}
\eeQ
for certain positive constants $\gamma$, $\delta$, where
\begin{align*}
\Phi(v)=\frac{\partial}{\partial v}\big(r \phi\big)(u_0, v)
\end{align*}
and $v \geq u\geq u_0$. The solution satisfies a priori estimates
 \begin{align*}
\partial _v r>\frac{1}{3}, \quad -\frac{1}{6} & > \partial _u r > -\frac{2}{3}, \quad \frac{2m}{r}<\frac{1}{2}\\
|\phi| \leq C \delta   \min\{1, r ^{-\gamma}\},  \quad |\partial _v &  (r \phi)|  \leq C \big(|\Phi(v)|+\delta \min\{1, r ^{-\gamma}\}\big), \\
|\partial _u (r \phi)| \leq C \delta, \quad
|\partial ^2 _v  (r \phi & ) | +|\partial ^2 _v r|+ |\partial ^2 _u (r \phi)| +|\partial ^2 _u r| \leq C
\end{align*}
where $C>0$ is some constant depending on $\gamma$. However, the solution does not have uniformly decaying estimates in $v$ unless the initial data satisfies
\beQ
\sup\limits_{v\in[u_0,\infty)}{\bigg\{(1+v)^{1+\ep} |\Phi(v)|+(1+v)^{2+\ep}|\pt_v\Phi(v)|\bigg\}}\leq A_0
\eeQ
further for some $A_0>0$ and $\ep>0$ \cite{LO, LOY}.

We point out that two metrics \eqref{me}, \eqref{double-null} are not equivalent in general. For instance, given $\Omega$, $r$ in \eqref{double-null}, in order to construct $g$ in \eqref{me}, we write \eqref{double-null} as
\begin{align*}
ds^2=-\Omega ^2 v_u   du ^2 -\Omega ^2 v_r dudr +r^2 \big(d\theta^2+\sin^2\theta d\phi^2 \big),
\end{align*}
and require
\begin{align*}
g \bar{g} =\Omega ^2 v_u, \,\, 2g=\Omega ^2 v_r \Longrightarrow \frac{4 v_u}{v_r}=\frac{1}{r}\int_0 ^r \Omega ^2 v_{r'} dr'.
\end{align*}
This does not hold true in general.

For some physical reason, cf. \cite{B} for instance, we refer solutions with the decaying property
\begin{align*}
h(0,r)=O(r^{-\epsilon}),\quad \frac{\partial h}{\partial r}(0,r)=O(r^{-1-\epsilon}),\quad as\,\,r\rightarrow \infty
\end{align*}
for some $\epsilon>0$ as particle-like decaying null infinity. On the other hand, in \cite{LZ}, Liu and Zhang studied global existence and uniqueness of classical solutions with small initial data, and of generalized solutions with large initial data for another type of decaying property. They proved that the solution satisfies the following uniformly decaying estimates
\beQ
|h(u,r)|\leq \frac{C}{(1+u+r)^{1+\ep}},\quad \Ba\frac{\pt h}{\pr}(u,r)\Ba\leq\frac{C}{(1+u+r)^{1+\ep}}
\eeQ
for $0<\ep \leq 2$, and the corresponding spacetime is future causally geodesically complete with vanishing final Bondi mass. That a function together with its derivative with respect to $r$ share the same decaying property as above is referred as wave-like decaying null infinity.

We refer to \cite{Chae, C02, WFAG} for the spherically symmetric Einstein-scalar field equations with nontrivial potential for particle-like decaying null infinity, and to \cite{LZ2} for wave-like decaying null infinity. We also refer to \cite{LR, LO1} for existence and uniqueness of Einstein fields equations coupled with scalar fields on non-spherically symmetric metrics for particle-like decaying spatial infinity. The slowly decaying of metrics at spatial infinity can be found in \cite{S1, S2} for vacuum Einstein field equations where the scalar field is zero.

In this paper, we prove global existence and uniqueness of classical solutions for the following small initial data sets, which we refer as slowly particle-like decaying null infinity.
\begin{thm}\label{M}
Let $0<\ep<1$. Given initial data $\breve{h}(r)\in C^1[0,\infty)$, denote
\begin{align*}
d_0 =\inf\limits_{b>0}\sup\limits_{r\geq 0}{\left\{\left (1+\frac{r}{b}\right)^{\ep} \left|\breve{h}(r)\right|
+\left(1+\frac{r}{b}\right)^{1+\ep}  \left|b \frac{\pt \breve{h} }{\pt r}(r)\right|\right\}}.
\end{align*}
Then there exists $\delta >0$ such that if $d_0 <\delta$, there exists unique global classical solution
\beQ
h(u,r) \in C^1 \Big([0,\infty)\times [0,\ift)\Big)
\eeQ
of (\ref{D}) which satisfies the initial condition $h(0,r)=\breve{h}(r)$ and the decay property
\beQ
|h(u,r)|\leq \frac{C}{\left(1+u+r\right)^{\ep}},\quad \left|\frac{\pt h}{\pr}(u,r)\right|\leq\frac{C}{\left(1+u+r\right)^{1+\ep}}
\eeQ
for some constant $C$ depending on $\ep$ only. Moreover, the corresponding spacetime is future causally geodesically complete. If we further assume
$\frac{1}{2}<\ep<1$, then the final Bondi mass vanishes.
\end{thm}

We adopt the main arguments in \cite{C1,LZ} to obtain the main estimates so as to prove the theorem. However, due to the slower decaying, we need to derive the estimates case by case much more carefully.

The paper is organized as follow: In Section 2, we derive the main estimates. In Section 3, we prove the main theorem.

\mysection{Main estimates}
\ls

In this section, we derive estimates of the iteration solutions as well as their partial derivatives with respect to $r$. Denote by $X$ the space of $C^1$ functions and the norm
\begin{align*}
\|h\|_{X}=\sup\limits_{u\geq 0, r\geq 0}{\left\{(1+u+r)^{\ep}|h(u,r)|+(1+u+r)^{1+\ep}\Big|\frac{\pt h}{\pt r}(u,r)\Big|\right\}}
\end{align*}
is finte. As in \cite{C1,LZ}, for $h_n\in X$, let $h_{n+1}$ be the solution of the equation
\beq
D_n h_{n+1}-\frac{g_n-\bar g_n}{2r}h_{n+1}=-\frac{g_n-\bar g_n}{2r}\bar h_n   \label{l-1}
\eeq
with the initial data
\beQ
h_{n+1}(0,r)=h(0,r),
\eeQ
where $g_n$ and $D_n$ are the metric given by \eqref{D} and the derivative along the incoming light rays with respect to metric $g_n$.
\begin{lem}\label{l1}
Let $0<\ep<1$. Given the initial data $h(0,r)$ and the nth-iteration solution $h_{n} (u,r)$ which are $C^1$ and satisfy
\begin{align*}
\|h(0,r)\|_{X}=d, \quad \|h_n(u,r)\|_X=x.
\end{align*}
Then the solution of (\ref{l-1}) satisfies
\begin{align*}
\|h_{n+1}\|_X \leq C(d+x^3)(2+x^2)\exp{(C x^2)} 
\end{align*}
for some positive constant $C$ independent of $n$.
\end{lem}
\pf By the assumption, we have
\beq
\begin{aligned}
|\bar h_n(u,r)|
&\leq \frac{1}{r}\int_{0}^{r}{|h_n(u,s)|ds}\\
&\leq \frac{x}{r}\int_{0}^{r}{\frac{ds}{(1+u+s)^{\ep}}}\\
&= \frac{x}{(1-\ep)r}\left[(1+u+r)^{1-\ep}-(1+u)^{1-\ep}\right]\\
&= \frac{x}{(1-\ep)r}\frac{(1+u+r)-(1+u)^{1-\ep}(1+u+r)^{\ep}}{(1+u+r)^{\ep}}\\
&\leq \frac{x}{(1-\ep)(1+u+r)^{\ep}}.
\end{aligned}\label{2-2}
\eeq
Using \eqref{2-2}, we can estimate $|(h_n-\bar h_n)(u,r)|$ as follows.

\noindent (1) For $0\leq r\leq 1+u$, same as \cite[(2.11)]{LZ}, we have
\beQ
\begin{aligned}
|(h_n-\bar h_n)(u,r)| \leq & \frac{1}{r}\ir{\int_{r'}^{r}{\left|\frac{\pt h_n}{\pt s}\right| ds}dr'}\\
\leq&\frac{x}{\ep r}\ir{\left[\frac{1}{\left(1+u+r'\right)^{\ep}}-\frac{1}{\ur^{\ep}}\right]dr'}\\
\leq& \frac{x}{\ep(1-\ep)r}\bigg[\ur^{1-\ep}-\uu^{1-\ep}\\
&-\frac{(1-\ep)r}{\ur^{\ep}}\bigg]\\
\leq &\frac{5xr(1+u)^{1-\ep}}{\ep(1-\ep)(1+u+r)^2}.\\
\end{aligned}
\eeQ
(2) For $r\geq 1+u$, we have
\beQ
\begin{aligned}
|(h_n-\bar h_n)(u,r)|&\leq |h_n(u,r)|+|\bar h_n(u,r)|\\
&\leq \frac{x}{\ur^{\ep}}+\frac{x}{1-\ep}\frac{1}{\ur^{\ep}}\\
&\leq \frac{2x}{1-\ep}\frac{(1+u+r)}{(1+u+r)^{1+\ep}}\\
&\leq \frac{4xr}{(1-\ep)\ur^{1+\ep}}.\\
\end{aligned}
\eeQ
Let $c=\frac{5}{\ep(1-\ep)}$. We obtain, for $r\geq 0$,
\beq
|(h_n-\bar h_n)(u,r)|\leq\left\{
\begin{aligned}
&\frac{cxr\uu^{1-\ep}}{\ur^2},\quad &0\leq r\leq 1+u,\\
&\frac{cxr}{\ur^{1+\ep}},\quad &r\geq 1+u.
\end{aligned}
\right.
\label{2-3}
\eeq
Thus
\beq
|(h_n-\bar h_n)(u,r)|\leq \frac{cxr}{\ur^{1+\ep}}. \label{2-4}
\eeq
Let $k=\exp{\left(-\frac{2\pi c^2x^2}{\ep}\right)}$, $0<k<1$. Since $g(u,r)$ is monotonically increasing with respect to $r$, we obtain
\beq
\begin{aligned}
\bar g_n(u,r)&\geq g_n(u,0)\\
&\geq \exp{\left[-4\pi\int_{0}^{\infty}{\frac{(h_n-\bar h_n)^2}{s}ds}\right]}\\
&\geq \exp{\left[-4\pi c^2x^2\int_{0}^{\infty}{\frac{ds}{(1+u+s)^{1+2\ep}}}\right]}\\
&\geq \exp{\left[-\frac{2\pi c^2x^2}{\ep(1+u)^{2\ep}}\right]}\geq k.\\
\end{aligned}\label{k}
\eeq

{\em Claim}: Let $c_1=\frac{4\pi c^2}{\ep}$. We have
\beq
\begin{matrix}
(g_n-\bar g_n)(u,r)\leq \left\{\begin{aligned}
&\frac{c_1x^2r^2}{(1+u)^{2\ep-1}\ur^3},\,\, &0\leq r\leq 1+u,\\
&\frac{c_1x^2r}{\uu^{2\ep}\ur},\,\,  &r\geq 1+u.\\
\end{aligned}\right.\label{2-5}
\end{matrix}
\eeq
Indeed, for $0\leq r\leq 1+u$, \eqref{2-5} is a direct consequence of (\ref{2-4}) and  \cite[(2.14)]{LZ}.
For $r\geq 1+u$, using (\ref{2-4}) and
\beQ
\frac{\pt g_n}{\pt r}=\frac{4\pi (h_n-\bar h_n)^2g_n}{r},
\eeQ
we obtain
\beQ
\begin{aligned}
(g_n-\bar g_n)(u,r)&=\frac{1}{r}\int_{0}^{r}{\int_{r'}^{r}{\frac{\pt g_n}{\pt s}ds}dr'}\\
&\leq \frac{2\pi c^2x^2}{\ep r}\int_{0}^{r}{\left[\frac{1}{(1+u+r')^{2\ep}}-\frac{1}{\ur^{2\ep}}\right]dr'}\\
&\leq \frac{2\pi c^2x^2}{\ep r}\left[\frac{r}{\uu^{2\ep}}-\frac{r}{\ur^{2\ep}}\right]\\
&=  \frac{2\pi c^2x^2}{\ep}\frac{(1+u+r)^2-\uu^{2\ep}\ur^{2-2\ep}}{\uu^{2\ep}\ur^2}\\
&\leq \frac{c_1 x^2r}{\uu^{2\ep}\ur}.
\end{aligned}
\eeQ
Thus, the claim follows and it also implies that
\beq
(g_n-\bar g_n)(u,r)\leq \frac{c_1 x^2r}{\uu^{2\ep+1}}.\label{g}
\eeq
Therefore,
\beQ
\begin{aligned}
\left|-\frac{g_n-\bar g_n}{2r}\bar h_n\right|&\leq \frac{1}{2r}\frac{c_1 x^2r}{\uu^{2\ep+1}}\frac{x}{1-\ep}\frac{1}{\ur^{\ep}}\\
&\leq \frac{c_1 x^3}{2(1-\ep)\uu^{2\ep+1}\ur^{\ep}}.
\end{aligned}
\eeQ

The characteristic $r(u)=\chi_n(u;r)$ satisfies
\begin{align*}
\frac{d\chi_n}{du}(u;r)=-\frac{1}{2}\bar{g_n}(u;r).
\end{align*}
Integrating it from $u$ to $u_1$, and using (\ref{k}), we obtain
\beQ
r(u)\geq r_1+\frac{1}{2}\int_{u}^{u_1}{\bar g_n du'}\geq r_1+\frac{k}{2}(u_1-u).
\eeQ
This implies that
\begin{align*}
1+u+r(u)&\geq \frac{k}{2}(1+u_1+r_1),\\
1+r_0&\geq \frac{k}{2}(1+u_1+r_1).
\end{align*}
They yield
\begin{align}
|h(0,r_0)|\leq & \frac{d}{(1+r_0)^{\ep}}\leq \frac{2^{\ep}d}{k^{\ep}(1+u_1+r_1)^{\ep}},   \label{2-7}\\
\int_{0}^{u_1}{\left[\frac{g_n-\bar g_n}{2r}\right]_{\chi_n}du}\leq &\frac{c_1x^2}{2}\int_{0}^{\infty}{\frac{du}{(1+u)^{2\ep+1}}}\leq \frac{c_1x^2}{4\ep},\label{2-8}
\end{align}
and
\beq
\begin{aligned}
\int_{0}^{u_1}&{\left|-\frac{1}{2r}(g_n-\bar g_n)\bar h_n\right|du}\\
\leq &\frac{c_1x^3}{2(1-\ep)}\int_{0}^{u_1}{\frac{du}{(1+u)^{2\ep+1}(1+u+r)^\ep}}\\
\leq &\frac{2^{\ep-1}c_1x^3}{k^{\ep}(1-\ep)\ur^{\ep}}\int_{0}^{u_1}{\frac{du}{(1+u)^{2\ep+1}}}\\
\leq & \frac{c_1x^3}{2^{2-\ep}k^{\ep}\ep(1-\ep)}\frac{1}{\ur^{\ep}}.\\
\end{aligned}\label{2-8a}
\eeq
Integrating (\ref{l-1}) along the characteristic $\chi_n$, we have
\beq
\begin{aligned}
h _{n+1}&(u_1, r_1)=h(0,r_0)\exp{\left\{\int_{0}^{u_1}{\left[\frac{g_n-\bgn}{2r}\right]_{\cn}du}\right\}}\\
&+\int_{0}^{u_1}\left[-\frac{g_n-\bgn}{2r}\bhn\right]_{\cn}
\exp\left\{\int_{u}^{u_1}{\left[\frac{g_n-\bgn}{2r}\right]_{\cn}du}\right\} du.
\end{aligned}\label{hn}
\eeq
Let $c_2=\frac{c_1}{2^{2-\ep}k^{\ep}\ep(1-\ep)}$. Substituting (\ref{2-7}), (\ref{2-8}), and (\ref{2-8a}) into (\ref{hn}), we obtain
\beq
(1+u_1+r_1)^{\ep} \left|h_{n+1}(u_1,r_1)\right|
\leq c_2\left(d+c_2x^3\right)\exp{\left(c_2x^2\right)}. \label{f}
\eeq

Next, we estimate $\frac{\partial h_{n+1}}{\partial r}$. Differentiate (\ref{l-1}) with respect to $r$, we obtain (cf. \cite[(9.16)]{C1} or \cite[(2.23)]{LZ})
\beq
D_n\frac{\pt h_{n+1}}{\pt r}-\frac{g_n-\bar g_n}{r}\frac{\pt h_{n+1}}{\pt r}=f_1,\label{G}
\eeq
where
\begin{align*}
f_1=&\frac{1}{2}\frac{\pt^2\bar g_n}{\pt r^2}(h_{n+1}-\bar h_n)-\frac{g_n-\bar g_n}{2r}\frac{\pt \bar h_n}{\pt r},\\
\frac{\pt^2\bar g_n}{\pt r^2}=&-\frac{2(g_n-\bar g_n)}{r^2}+\frac{4\pi (h_n-\bar h_n)^2}{r^2}g_n.
\end{align*}
Integrating (\ref{G}) along the characteristic $\chi_n$, we obtain
\beq
\begin{aligned}
\frac{\pt h_{n+1}}{\pt r}&(u_1,r_1)=\frac{\pt h}{\pt r}(0,r_0)\exp{\left\{\int_{0}^{u_1}{\left[\frac{g_n-\bar g_n}{r}\right]_{\chi_n}du}\right\}} \\
&+\int_{0}^{u_1}{\exp{\left\{\int_{u}^{u_1}{\left[\frac{g_n-\bar g_n}{r}\right]_{\chi_n}du'}\right\}}[f_1]_{\chi_n}du}.
\end{aligned}
\label{phn}
\eeq
Using (\ref{2-3}) and (\ref{2-5}), we have, for $0 \leq r \leq 1+u$,
\begin{align*}
\left|\frac{\pt^2\bar g_n}{\pt r^2}\right|
\leq \frac{2}{r^2}\frac{c_1x^2r^2}{\uu^{2\ep-1}\ur^3}+\frac{4\pi}{r^2}
\frac{c^2x^2r^2\uu^{2-2\ep}}{\ur^4}
\end{align*}
and, for $r \geq 1+u$,
\begin{align*}
\left|\frac{\pt^2\bar g_n}{\pt r^2}\right|
\leq
\frac{2}{r^2}\frac{c_1x^2r}{\uu^{2\ep}\ur}+\frac{4\pi}{r^2}\frac{c^2x^2r^2}{\ur^{2+2\ep}}.
\end{align*}
Therefore
\begin{align}
\left|\frac{\pt^2\bar g_n}{\pt r^2}\right|\leq \frac{(2c_1+4\pi c^2)x^2}{\uu^{2\ep+1}\ur}.  \label{bg}
\end{align}
Let $c_3=\frac{c_2(2\pi c^2+c_1)}{1-\ep}$, $c_4=c_3+\frac{cc_1}{2}$. Using (\ref{2-2}), (\ref{2-4}), (\ref{g}), and (\ref{f}), we obtain
\begin{align*}
\frac{1}{2}\left|\frac{\pt^2\bar g_n}{\pt r^2}\right|\cdot|h_{n+1}-\bar h_n|
&\leq \frac{c_3x^2(d+x+x^3)\exp{(c_2x^2)}}{\uu^{2\ep+1}\ur^{1+\ep}},\\
\left|\frac{g_n-\bar g_n}{2r}\frac{\pt \bar h_n}{\pt r}\right|&=\left|\frac{(g_n-\bar g_n)(h_n-\bar h_n)}{2r^2}\right|\\
&\leq \frac{1}{2r^2}\frac{c_1x^2r}{\uu^{2\ep+1}}\frac{cxr}{\ur^{1+\ep}}\\
&\leq \frac{cc_1 x^3}{2\uu^{2\ep+1}\ur^{1+\ep}}.
\end{align*}
Therefore,
\begin{align}
|f_1|\leq \frac{c_4x^2(d+x+x^3)\exp{(c_2x^2)}}{\uu^{2\ep+1}\ur^{1+\ep}}.
\label{2-14}
\end{align}
Similar to (\ref{2-7}), we have
\begin{align}
\left|\frac{\pt h}{\pt r}(0,r_0)\right|\leq \frac{2^{1+\ep}d}{k^{1+\ep}(1+u_1+r_1)^{1+\ep}}.\label{2-15}
\end{align}
Let $c_5=\max\big\{2c_2,\frac{2^{\ep}c_4}{\ep k^{1+\ep}}\big\}$. Using (\ref{2-8}), (\ref{2-14}), and (\ref{2-15}), (\ref{phn}) yields
\beq
\left|\frac{\pt h_{n+1}}{\pt r}(u_1,r_1)\right|\leq \frac{c_5(d+x^3)(1+x^2)\exp{(c_5x^2)}}{(1+u_1+r_1)^{1+\ep}}.
\label{2-16}
\eeq
Let $C=2\max{\{c_5,\, c_2^2\}}$. (\ref{f}) and (\ref{2-16}) imply that
\begin{align*}
\|h_{n+1}\|_X \leq C(d+x^3)(2+x^2)\exp{(C x^2)}.
\end{align*}
Thus proof of the lemma is complete. \qed

\mysection{Proof of the main theorem}
\ls

In this section we prove Theorem \ref{M}. Denote $\{h_n\}$ the sequence of the iteration solutions of \eqref{l-1}.
We first show that $\{h_n\}$ converges in function space
\beQ
Y=\left\{h\in C^0 [0,\infty)\times[0,\infty) \Big | \|h\|_Y <\infty     \right\},
\eeQ
where
\beQ
\|h\|_Y=\sup\limits_{u\geq 0, r\geq 0 }{\Big \{(1+u+r)^{\ep}|h(u,r)|\Big \}}.
\eeQ

\begin{lem}\label{l2}
Assume there exists some $x>0$ such that
\begin{align*}
\|h_{n-1}\|_X\leq x, \quad \|h_n\|_X\leq x,
\end{align*}
then there exists $F(x)\in(0,\frac{1}{2})$ such that
\beQ
\|h_{n+1}-h_n \|_Y\leq F(x)\|h_n-h_{n-1}\|_Y.
\eeQ
\end{lem}
\pf Taking the derivative $D_n$ along the characteristic $\chi_n$, we obtain (cf. \cite[(9.27)]{C1})
\beq
D_n(h_{n+1}-h_n)-\frac{g_n-\bar g_n}{2r}(h_{n+1}-h_n) =f_2,\label{3-1}
\eeq
where
\beQ
\begin{aligned}
f_2=&\frac{\bar g_n-\bar g_{n-1}}{2}\frac{\pt h_n}{\pt r}-\frac{g_n-\bar g_n}{2r}(\bar h_n-\bar h_{n-1})\\
&+\frac{g_n-\bar g_n-g_{n-1}+\bar g_{n-1}}{2r}(h_n-\bar h_{n-1}).
\end{aligned}
\eeQ
Similar to (\ref{2-2}), we have
\beQ
\big |\bar h_n-\bar h_{n-1} \big |\leq \frac{\hy}{(1-\ep)\ur^{\ep}}.
\eeQ
Then
\beq
\big |h_n-h_{n-1}-\bar h_n+\bar h_{n-1} \big |\leq \frac{2\hy}{(1-\ep)\ur^{\ep}}.\label{3-3}
\eeq
From (\ref{2-3}), we obtain
\beq
\begin{matrix}
\big |h_n+h_{n-1}-\bar h_n-\bar h_{n-1} \big |\leq
\left\{\begin{aligned}
&\frac{2cxr\uu^{1-\ep}}{\ur^2},\,&0\leq r\leq 1+u\\
&\frac{2cxr}{\ur^{1+\ep}},\,&r\geq 1+u.\\
\end{aligned}
\right.
\end{matrix}\label{3-4}
\eeq
Thus, (\ref{3-3}) and (\ref{3-4}) yield
\beq
\begin{aligned}
\big |(h_n- &\bar h_n)^2-(h_{n-1}-\bar h_{n-1})^2 \big| \\
\leq &\left\{\begin{aligned}
&\frac{4cxr\uu^{1-\ep}\hy}{(1-\ep)\ur^{2+\ep}},\,&0\leq r\leq 1+u\\
&\frac{4cxr\hy}{(1-\ep)\ur^{1+2\ep}},\,&r\geq 1+u.\\
\end{aligned}
\right.
\end{aligned}\label{3-5}
\eeq
This implies that
\beq
\begin{aligned}
|g_n-g_{n-1}|&\leq 4\pi\int_{r}^{\infty}{\Big|(h_n-\bar h_n)^2-(h_{n-1}-\bar h_{n-1})^2\Big|\frac{ds}{s}}  \\
&\leq \frac{16\pi cx \hy}{\ep(1-\ep)\ur^{2\ep}}.\label{3-6}
\end{aligned}
\eeq
Therefore,
\begin{align}
|\bar g_n-\bar g_{n-1}| \leq \frac{1}{r}\int_{0}^{r}{|g_n-g_{n-1}|dr}\leq \frac{16\pi cx \hy }{\ep(1-\ep)(1+u)^{2\ep}}. \label{3-7}
\end{align}
Let $c_6=\frac{8\pi cc_5}{\ep(1-\ep)}$. Using \eqref{2-16} and (\ref{3-7}), we obtain
\begin{align}
\left|\frac{\bar g_n-\bar g_{n-1}}{2}\frac{\pt h_n}{\pt r}\right|
\leq \frac{c_6(d+x^3)(x+x^3)\hy\exp{(c_5x^2)}}{\uu^{2\ep+1}\ur^{\ep}}.  \label{3-8}
\end{align}
Using (\ref{2-2}) and (\ref{g}), we obtain
\begin{align}
\Big |\frac{g_n -\bar g_n}{2r}(\bar h_n-\bar h_{n-1})\Big |
\leq  \frac{c_1x^2 \hy}{(1-\ep) \uu^{2\ep+1}\ur^{\ep}}.  \label{3-9}
\end{align}
Let $c_7=\frac{16\pi^2c^3}{\ep^2(1-\ep)}$. (\ref{2-4}), (\ref{3-5}), and (\ref{3-6}) yield
\begin{align*}
\frac{1}{2r} \Big|g_n -&g_{n-1} -(\bar g_n-\bar g_{n-1})\Big|\\
\leq & \frac{1}{2r^2} \int_{0}^{r} {\int_{r'}^{r}{\left|\frac{\pt(g_n-g_{n-1})}{\pt s}\right| ds}dr'}\\
\leq & \frac{2\pi}{r^2} \int_{0}^{r} \int_{r'} ^{r} |g_n| \big|(h_n-\bar h_n)^2-(h_{n-1}-\bar h_{n-1})^2 \big |  \frac{ds}{s} dr'\\
     & +\frac{2\pi}{r^2} \int_{0}^{r} \int_{r'}^{r} |g_n-g_{n-1}| |h_{n-1}-\bar h_{n-1}|^2 \frac{ds}{s}  dr'\\
\leq &\frac{32\pi^2 c^3(x+x^3)\hy}{\ep(1-\ep)r^2}\int_{0}^{r}{\int_{r'}^{r}{\frac{dsdr'}{(1+u+s)^{1+2\ep}}}}\\
\leq &\frac{c_7 (x+x^3)\hy }{r^2}\int_{0}^{r}{\left[\frac{1}{\uu^{2\ep}}-\frac{1}{\ur^{2\ep}}\right]dr'}\\
\leq &\frac{2c_7(x+x^3)\hy}{\uu^{2\ep+1}}.
\end{align*}
Thus, together with (\ref{2-2}) and (\ref{f}), we obtain
\beq
\begin{aligned}
\frac{1}{2r}&\Big|g_n-g_{n-1}-(\bar g_n-\bar g_{n-1})\Big| \left|h_n-\bar h_{n-1}\right|\\
&\leq \frac{2c_2^2c_7(d+x^3)(x+x^3)\exp{(c_2x^2)}}{\uu^{1+2\ep}\ur^{\ep}}.
\end{aligned}\label{3-11}
\eeq
Let $c_8=c_6+\frac{c_1}{1-\ep}+2c_2^2c_7$. Using (\ref{3-8}), (\ref{3-9}), and (\ref{3-11}), we obtain
\beq
|f_2|\leq \frac{c_8(d+x+x^3)(x+x^3)\hy\exp{(c_8x^2)}}{\uu^{2\ep+1}\ur^{\ep}}.
\label{3-12}
\eeq
Integrating (\ref{3-1}) along the characteristic $\chi_n$, and using (\ref{2-8}) and (\ref{3-12}), we have
\beQ
(1+u_1+r_1)^{\ep}\big|(h_{n+1}-h_n)(u_1,r_1) \big |\leq F(x)\hy,
\eeQ
where
\beQ
F(x)=\frac{2^\ep c_8(d+x+x^3)(x+x^3)\exp{(2c_8x^2)}}{2\ep k^{\ep}}.
\eeQ
Obviously,
\beQ
F(0)=0,\quad F'(x)\geq 0.
\eeQ
Thus $F(x)$ is monotonically increasing. Therefore there exists $x_1 >0$ such that, for any $x \in (0,x_1)$,
\beQ
0< F(x)< \frac{1}{2}.
\eeQ
This gives proof of the lemma.\qed

Next, we show that the sequences $\{h_n\}$ and $\{\frac{\pt h_n}{\pt r}\}$ are uniformly bounded and equicontinuous.
\begin{lem}\label{l3}
There exists $\tilde x$ such that for any $x\in(0,\tilde x)$, the sequence $\{h_n\}$ are uniformly bounded by $x$ in the space $X$ and converges in the space $Y$ if
\begin{align*}
\Phi(x)=\frac{x\exp\left(-Cx^2\right)}{C(2+x^2)}-x^3 \geq d,
\end{align*}
where $d=\|h(0,r)\|_X$ (see Lemma \ref{l1}). Moreover, $\{h_n\}$ and $\{\frac{\pt h_n}{\pt r}\}$ are uniformly bounded and equicontinous.
\end{lem}
\pf By Lemma \ref{l1}, we know that
\beQ
\|h_{n+1}\|_X\leq C(d+x^3)(2+x^2)\exp{(C x^2)}.
\eeQ
Denote
\beQ
\Phi(x)=\frac{x\exp\left(-C x^2\right)}{C (2+x^2)}-x^3.
\eeQ
It is clearly that
\beQ
\Phi(0)=0,\quad \Phi'(0)>0.
\eeQ
Thus, there exists $x_0$ such that $\Phi(x)$ is monotonically increasing on $[0,x_0]$ and attains its maximum at $x_0$. Let
\beQ
\tilde x=\min{\{x_0,x_1,1\}},
\eeQ
where $x_1$ is given in Lemma \ref{l2}. Then for any $x\in (0,\tilde x)$,
\begin{align*}
\Phi(x) \geq d, \quad \|h_n\|_X\leq x \Longrightarrow  \|h_{n+1}\|_X\leq x.
\end{align*}
By induction, $\|h_n\|\leq x$ for all $n\in\mathbb{N}$, i.e., $\{h_n\}$ is uniformly bounded by $x$. By the proof of Lemma 2.3 in \cite{LZ}, we have
\begin{align*}
\left|\frac{\pt h_n}{\pt r}\right|&\leq x,\\
\left|\frac{\pt h_n}{\pt u}\right|&\leq \frac{\bar g_{n-1}}{2}\left|\frac{\pt h_n}{\pt r}\right|+\frac{1}{2r}(g_{n-1}-\bar g_{n-1})|h_{n-1}-\bar h_{n-1}|\leq \frac{1}{2}(x+cc_1x^3).
\end{align*}
These imply that both $\{\frac{\pt h_n}{\pt r}\}$ and $\{\frac{\pt h_n}{\pt u}\}$ are uniformly bounded. Thus, $\{h_n\}$ is equicontinuous.

By Lemma \ref{l2}, we obtain that for any $x\in (0,\tilde x)$,
\beQ
\|h_{n+1}-h_n\|_Y< \frac{1}{2}\|h_n-h_{n-1}\|_Y.
\eeQ
This implies that $\{h_n\}$ converges in space $Y$.

For any $u \geq 0$, $0\leq r_1<r_2$, let $\chi_n(u;r_1)$ and $\chi_n(u;r_2)$ be two characteristics through $u$-slice at $r=r_1$ and $r=r_2$ respectively. Let $k'=\exp{\left(\frac{c_1x^2}{4\ep}\right)}$. By \cite[(4.29)]{C2} and (\ref{g}), we obtain
\beQ
\begin{aligned}
\left|\frac{\chi_n(u;r_2)-\chi_n(u;r_1)}{r_2-r_1} \right|
& \leq
\sup\limits_{s\in [r_1,r_2]}{\exp{\left\{\frac{1}{2}\int_{u}^{u_1}{\left[\frac{\pt\bar g}{\pt r}\right]_{\chi_n(u';s)}du'}\right\}}} \\
& \leq k' . 
\end{aligned}
\eeQ

For any differentiable function $f$, define
\beQ
B(f)(u)=f(u,\cn(u;r_1))-f(u,\cn(u;r_2)).
\eeQ
We have
\beQ
|B(f)(u)|\leq \sup{\left|\frac{\pt f}{\pt r}\right|}k'(r_2-r_1).
\eeQ
Now we use the argument in \cite{LZ} to prove that $\{\frac{\pt h_n}{\pt r}\}$ is equicontinuous. Let
\beq
\psi(u)=\frac{\pt h_{n+1}}{\pr}(u,\cn(u;r_1))-\frac{\pt h_{n+1}}{\pr}(u,\cn(u;r_2)). \label{psi}
\eeq
Differentiate (\ref{psi}), we have
\beq
\psi'(u)-\frac{(g_n-\bgn)(u,\cn(u;r_1))}{\cn(u;r_1)}\psi(u)=\sum _{i=1} ^4 A_i,\label{psiU}
\eeq
where
\beQ
\begin{aligned}
A_1&=\frac{\pt h_{n+1}}{\pr}(u,\cn(u;r_2))B\Bl\frac{g_n-\bgn}{r}\Br(u),\\
A_2&=\frac{1}{2}B\Bl\frac{\pt^2\bgn}{\pt r^2}(h_{n+1}-\bar{h}_{n+1})\Br(u),\\
A_3&=\frac{1}{2}B\Bl\frac{\pt^2\bgn}{\pt r^2}(\bar{h}_{n+1}-\bhn)\Br(u),\\
A_4&=-\frac{1}{2}B\Bl\frac{\pt\bgn}{\pr}\frac{h_n-\bhn}{r}\Br(u).
\end{aligned}
\eeQ
From (\ref{bg}), we have
\beQ
\begin{aligned}
|A_1|&\leq \left|\frac{\pt h_{n+1}}{\pt r}\right|\cdot\left|\frac{\pt^2 \bar g_n}{\pt r^2}\right|k'(r_2-r_1)\\
&\leq k'(r_2-r_1)\frac{x}{\uu^{1+\ep}}\frac{(2c_1+4\pi c^2)x^2}{\uu^{2\ep+1}\ur}\\
&\leq \frac{k'(2c_1+4\pi c^2)x^3}{\uu^{3+3\ep}}(r_2-r_1).\\
\end{aligned}
\eeQ
Next, to estimate $A_2$, we need to derive the estimate of the 3rd partial derivative of $\bar g_n$ with respect to $r$. By direct calculation, we obtain (cf. \cite[(2.42)]{LZ})
\begin{align*}
\frac{\partial ^3 \bar g_n}{\partial r^3}=&\frac{6(g_n -\bar g _n)}{r^3}-\frac{16\pi (h_n -\bar h _n)^2  g_n}{r^3} \\
& +\frac{8\pi (h_n -\bar h _n)g_n }{r^2}\frac{\partial (h_n -\bar h _n) }{\partial r}+\frac{16\pi ^2  (h_n -\bar h _n )^4 g_n}{r^3}.
\end{align*}
Using (\ref{2-3}) and (\ref{g}), we obtain
\beq
\left|\frac{\partial ^3 \bar g_n}{\partial r^3}\right|\leq
\left\{\begin{aligned}
&\frac{(6c_1+32\pi c^2)x^2+16\pi^2c^4x^4}{r\uu^{2\ep+1}},&0\leq r\leq 1+u,\\
&\frac{(6c_1+32\pi c^2)x^2+16\pi^2c^4x^4}{\uu^{2\ep+2}},&r\geq 1+u,\\
\end{aligned}
\right.\label{bg3}
\eeq
Let $c_9=10cc_1+40\pi c^3+16\pi^2 c^5.$ Then (\ref{2-3}) and (\ref{bg3}) imply that
\beQ
|A_2|\leq \frac{k'c_9(x^3+x^5)}{\uu^{3\ep+2}}(r_2-r_1).
\eeQ
By Lemma \ref{l2}, we know
\beQ
h_{n+1}-h_n\rightarrow 0
\eeQ
uniformly. Then the argument for proving \cite[Lemma 2.3]{LZ} gives
\beQ
\bar h_{n+1}-\bar h_n\rightarrow 0.
\eeQ
By (\ref{bg}), we have
\beQ
\left|(1+u)^{2\ep+1}\frac{\pt^2\bar g_n}{\pt r^2}\right|\leq (2c_1+4\pi c^2)x^2.
\eeQ
Thus,
\beQ
(1+u)^{2\ep+1}\frac{\pt^2\bar g_n}{\pt r^2}(\bar h_{n+1}-\bar h_n)\rightarrow 0
\eeQ
uniformly. Therefore,
\begin{align*}
(1+u)^{2\ep+1}\frac{\pt^2\bar g_n}{\pt r^2}(\bar h_{n+1}-\bar h_n)
\end{align*}
is equicontinuous. Hence for $\eta>0$, there exists $t>0$ such that
\begin{align*}
|\cn(u;r_2)-\cn(u;r_1)|\leq t \Longrightarrow |A_3|\leq \frac{2\ep \eta}{3k'^2\uu^{2\ep+1}}.
\end{align*}
Taking $s_1=\frac{t}{k'}$, we have
\begin{align*}
r_2-r_1\leq s_1 & \Longrightarrow |\cn(u;r_2)-\cn(u;r_1)|\leq t \\
                & \Longrightarrow |A_3|\leq \frac{2\ep\eta}{3k'^2\uu^{2\ep+1}}.
\end{align*}
Similarly,
\beQ
|A_4|\leq \frac{k'(3cc_1+2\pi c^4)x^3}{\uu^{3+3\ep}}(r_2-r_1).
\eeQ
Taking
\begin{align*}
s_2=\frac{2\ep\eta}{3k'^3[(2c_1+4\pi c^2+c_9+3cc_1+2\pi c^4)x^3+c_9x^5]},
\end{align*}
we have
\begin{align*}
r_2-r_1\leq s_2 \Longrightarrow |A_1|+|A_2|+|A_4|\leq \frac{2\ep\eta}{3k'^2\uu^{2\ep+1}}.
\end{align*}
By (\ref{g}), we have
\begin{align*}
\exp{\Bl\int_{0}^{u_1}{\frac{(g_n-\bgn)(u,\cn(u;r_1))}{\cn(u;r_1)}du}\Br}\leq k'^2.
\end{align*}
Since $\frac{\pt h}{\pt r}(0,r)$ is uniformly continuous, there exists $t'>0$ such that (cf. \cite[Lemma 2.3]{LZ})
\begin{align*}
|\cn(0;r_1)&-\cn(0;r_2)|\leq t' \\
&\Longrightarrow \left|\frac{\pt h}{\pr}(0,\cn(0;r_1))-\frac{\pt h}{\pr}(0,\cn(0;r_2))\right|\leq \frac{\eta}{3k'^2}.
\end{align*}
Thus, taking $s_3=\frac{t'}{k'}$, we have
\begin{align*}
r_2-r_1\leq s_3&\Longrightarrow |\cn(0;r_1)-\cn(0;r_2)|\leq t'\\
&\Longrightarrow \left|\frac{\pt h}{\pr}(0,\cn(0;r_1))-\frac{\pt h}{\pr}(0,\cn(0;r_2))\right|\leq \frac{\eta}{3k'^2}.
\end{align*}
Therefore, integrating (\ref{psiU}), we obtain
\beQ
\begin{aligned}
\psi(u_1)=&\psi(0)\exp{\Bl\int_{0}^{u_1}{\frac{(g_n-\bgn)(u,\cn(u;r_1))}{\cn(u;r_1)}du}\Br}\\
          &+\int_{0}^{u_1}{\Big[\exp{\Bl\int_{u}^{u_1}{\frac{(g_n-\bgn)(u,\cn(u;r_1))}{\cn(u;r_1)}du}\Br}\Big]\sum _{i=1} ^4 A_i du}.
\end{aligned}
\eeQ
Let $s=\min{\{s_1,s_2,s_3\}}$. Then
\begin{align*}
r_2-r_1\leq s \Longrightarrow |\psi(u_1)|\leq \eta \Longleftrightarrow
\left|\frac{\pt h_{n+1}}{\pr}(u_1,r_1)-\frac{\pt h_{n+1}}{\pr}(u_1,r_2)\right|\leq\eta.
\end{align*}
Thus, $\{\frac{\pt h_{n+1}}{\pt r}\}$ is equicontinuous with respect to $r$.

The equicontinuous of $\{\frac{\pt h_{n+1}}{\pt r}\}$ with respect to $u$ can be proved by the equiboundedness of $D_n \frac{\partial h_{n+1}}{\partial r}$. Thus proof of the lemma is complete. \qed

{\em Proof of Theorem \ref{M}.} Denote
\begin{align*}
\delta=\max\limits_{[0,\tilde x]}{\{\Phi(x)\}}.
\end{align*}
For $d_0\leq \delta$, we can find $x$ such that $d_0 \leq \Phi(x)$. This implies that Lemma \ref{l1}, Lemma \ref{l2} and Lemma \ref{l3} hold. Given
initial data $\breve{h}(r)$ with $d_0\leq \delta $, then there exists $a>0$ such that
\begin{align*}
\hat{d}_0 =\sup\limits_{r\geq 0}{\left\{\left (1+\frac{r}{a}\right)^{\ep} \left|\breve{h}(r)\right|
+\left(1+\frac{r}{a}\right)^{1+\ep}  \left|a \frac{\pt \breve{h} }{\pt r}(r)\right|\right\}}<\delta.
\end{align*}
Consider the new initial data
\beQ
\hat h(0,r)=\breve{h}(ar).
\eeQ
By using the same argument as these in \cite{C1, LZ}, there exists a unique global classical solution $\hat h(u,r)$ with slowly particle-like decaying null infinity satisfying the initial data $\hat h(0,r)$. By scaling group invariance of (\ref{l-1}) (see \cite{C1}), we find that
\beQ
h(u,r)=\hat h\left(\frac{u}{a},\frac{r}{a}\right)
\eeQ
is a unique global classical solution of (\ref{l-1}) satisfying the initial data $\breve{h}(r)$. Moreover $h$ satisfies
\begin{align*}
|h(u,r)|\leq \frac{C}{(1+u+r)^{\ep}},\quad \left|\frac{\pt h}{\pt r}(u,r)\right|\leq \frac{C}{(1+u+r)^{1+\ep}}.
\end{align*}
Now (\ref{2-3}) and (\ref{k}) imply that
\beQ
k\leq g,\,\bar g\leq 1
\eeQ
By Lemma \ref{l1}, $h$, $\bar h$, $g$, $\bar g$ and their partial derivatives are all uniformly bounded. Thus, using the same argument as \cite{LZ}, we can show that the corresponding spacetime is future casually geodecsically complete.

Finally, using the same argument as \cite{LZ}, we have
\beQ
\begin{aligned}
1-g(u,r)&\leq 4\pi \int_{r}^{\infty}{\frac{(h-\bar h)^2}{r'}dr'}\\
&\leq 4\pi c^2x^2\int_{r}^{\infty}{\frac{r'}{(1+u+r')^{2+2\ep}}dr'}\\
&\leq \frac{2\pi c^2x^2}{\ep(1+u+r)^{2\ep}}.
\end{aligned}
\eeQ
Therefore
\beQ
\frac{r}{2} \Big (1-g \Big)\leq \frac{\pi c^2x^2}{\ep}\frac{r}{(1+u+r)^{2\ep}}.
\eeQ
We obtain
\begin{align*}
\frac{1}{2}<\ep<1   \Longrightarrow\lim\limits_{r\rw\infty}{\frac{r}{2}\Big(1-g\Big)}=0.
\end{align*}
Hence the Bondi-Christodoulou mass is equivalent to the Bondi mass
\beQ
M(u)=M_B(u).
\eeQ
From \cite{C1}, we have
\beQ
\begin{aligned}
M(u)=2\pi \int_{0}^{\infty}{\frac{\bar g}{g}(h-\bar h)^2dr}\leq \frac{2\pi c^2x^2}{2\ep-1}\frac{1}{(1+u)^{2\ep-1}}.
\end{aligned}
\eeQ
Therefore,
\begin{align*}
\frac{1}{2}<\ep<1   \Longrightarrow M_1=\lim\limits_{u\rw\infty}{M(u)}=0.
\end{align*}
Thus proof of the theorem is complete. \qed

\bigskip

\footnotesize {

\noindent {\bf Acknowledgement} The work is supported by the National Natural Science Foundation of China (Nos. 12301072, 12326602).

\noindent {\bf Conflict of Interest} The authors declare no conflict of interest.

}

\end{document}